\documentclass{aa}
\usepackage{txfonts} 
\usepackage{graphicx}   

\begin{document}

\def\l{$\lambda$}
\def\ie{{\it i.e.}}
\def\kms{~km~s$^{-1}$}
\def\micron{$\mu$m}
\def\fe2{[\ion{Fe}{ii}]}
\def\h2{H$_{2}$}

\title {Optical and NIR spectroscopy of Mrk~1210: constraints 
and physical conditions of the active nucleus}

\author{
Ximena Mazzalay\inst{1}\fnmsep\thanks{e-mail: ximena@oac.uncor.edu}
\and
Alberto Rodr\'{\i}guez-Ardila\inst{2}\fnmsep\thanks{Visiting Astronomer at the Infrared Telescope Facility, which is operated by the University of Hawaii under Cooperative Agreement no. NCC 5-538 with the National Aeronautics and Space Administration, Office of Space Science, Planetary Astronomy Program.}}

\institute{IATE, Observatorio Astron\'omico C\'ordoba, Laprida 854, X5000BGR, 
C\'ordoba, Argentina.
\and
Laborat\'orio Nacional de Astrof\'{\i}sica, Rua dos Estados Unidos 154, Itajub\'a, MG, Brazil
}

\date{Received; --- accept}

\abstract{Near-infrared and optical spectroscopy of the nuclear and 
extended emission region of the Seyfert~2 galaxy Mrk~1210 is presented. 
This galaxy is outstanding because it displays signatures of recent 
circumnuclear star formation and a high-level of X-ray activity, 
in addition to the classical spectral characteristics typical of an AGN. 
The NIR nuclear spectrum, which covers the interval 
0.8--2.4~$\mu$m, is dominated by \ion{H}{i} and \ion{He}{i} 
recombination lines as well as
[\ion{S}{ii}], [\ion{S}{iii}] and \fe2 forbidden lines. 
Coronal lines of [\ion{S}{viii}], 
[\ion{S}{ix}], [\ion{Si}{vi}], [\ion{Si}{x}], and [\ion{Ca}{viii}]
in addition to molecular H$_{2}$ lines are also detected.
Outside the nuclear region, extended emission of [\ion{S}{iii}] and
\ion{He}{i} is found up to a distance of 500~pc from the center.
An estimate of the contribution of the stellar population
to the continuum is made by means of the $^{12}{\rm CO(6-3)}$~1.618~$\mu$m
overtone bandhead. It was found that $83\pm8\%$ of the H-band continuum
is of stellar origin. It improves previous estimates, which 
claimed that at least 50\% of the observed continuum was attributed
to the AGN. The analysis of the emission
line profiles, both allowed and forbidden, shows a narrower  
(${\rm FWHM} \sim 500$~\kms) line on top of a broad (${\rm FWHM} > 1\,000$~\kms) 
blue-shifted component. The latter seems to be associated to 
a nuclear outflow. This hypothesis is supported by 6~cm VLBI 
observations, which show a radio ejecta extending of up to $\sim$30~pc 
from the nucleus. This result does not required the presence of a hidden 
BLR claimed to be present in previous NIR observations 
of this object. Internal extinction, calculated by
means of several indicators including [\ion{Fe}{ii}] flux
ratios not previously used before in AGNs, reveals a dusty AGN,
while the extended regions are little or not affected by dust. 
Density and temperature for the NLR are calculated using optical and
NIR lines diagnostic ratios. The results show electronic temperatures 
from 10\,000~K up to 40\,000~K and densities between $10^3 - 10^5$~cm$^{-3}$.
The larger temperatures points out that shocks, most probably
related to the radio outflow, must contribute 
to the line emission.

}

\authorrunning{Mazzalay \& Rodr\'{\i}guez-Ardila}
\titlerunning{Optical and NIR spectroscopy of Mrk~1210}

\maketitle


\section{Introduction}
Mrk~1210, also known as the Phoenix Galaxy, is a nearby Sa Seyfert~2 galaxy 
at $z=0.01350$. It has been extensively studied in the optical region,
mainly because of the Wolf-Rayet features found within the central 200~pc 
(Storchi-Bergmann et al. 1998), indicating the 
presence of a circumnuclear starburst, and the 
detection of broad H$\alpha$ and H$\beta$
components in polarized light (Tran et al. 1992; Tran 1995). The widths of
these last components reach 2380~km~s$^{-1}$ at FWHM. Near-infrared (NIR) 
spectra reported by Veilleux et al. (1997) 
show that the Pa$\beta$ profile is characterized by a strong narrow component on top
of a broad base with ${\rm FWHM} \sim 1600$~km~s$^{-1}$, suggesting the presence of 
a hidden broad line region (BLR).  
However, a comparison of the Pa$\beta$ profile with that of \fe2~1.256~\micron\ 
indicates that at least part of the broad emission is also present in the forbidden \fe2\ line.
The moderate critical densities of the NIR \fe2\ lines, $n_\mathrm{c} \sim 10^{4-5}$~cm$^{-3}$ 
(Nisini et al. 2002), and the broad emission observed in \fe2~1.256~\micron\ suggest 
that the broad permitted line is not 
produced in a genuine high-density BLR. 
The possible
presence of a broad component to Pa$\beta$ and Br$\gamma$ and the absence
of broad H$\alpha$ in direct light would imply an 
${\rm E(B-V)_b} >$ 1.6 and 1.1,
respectively, meaning that this object is strongly reddened.

Additional evidence of the dusty nature of Mrk~1210 is abundant. It is a member 
of the sample of galaxies called 60~$\mu$m
peakers (60PKs, Heisler \& De Roberties 1999) because of its ``warm''
far-infrared color and the spectral energy distribution peaking 
near 60~$\mu$m. Over the past few years, evidence has been
accumulating indicating that these properties can be attributed to dust-obscured
active galactic nucleus (e.g., Keel et al. 1994; Hes et al. 1995;
Heisler \& Vader, 1995). Whether or not this obscured
material is the putative dusty torus as postulated by the
unified model (e.g., Antonucci 1993) is highly debatable 
although evidence from spectropolarimetry points out to this scenario. 
In fact, Heisler et al. (1997) showed that Seyfert~2
galaxies for which polarized broad lines have been detected have 
warmer far-infrared colors ($f_{60}/f_{25} < 4$) compared to
Seyfert~2 galaxies for which polarized broad lines have not been detected. 
The warm IRAS Seyferts are believed to be viewed at an angle that is more
face-on than that of cool IRAS Seyferts. Moreover, in the near-infrared, the 
degree of polarization of Mrk~1210 rises toward longer wavelengths, 
reaching $\sim 5$\% at 2.15~$\mu$m (Watanabe et al. 2003). Observational 
evidence suggest that this polarization is due to dichroic absorption by 
aligned grains. Watanabe et al. (2003) also suggest the existence of 
an additional unpolarized component, most 
probably associated to electron scattering.

The picture of a highly reddened nucleus derived from polarimetry
is also strongly supported by optical multicolor imaging. Indeed, 
Heisler \& Vader (1994) 
showed that Mrk~1210 displays a smooth global light distribution
but morphologically resembles a typical elliptical galaxy. Masked frames
display arcs or shells surrounding the main body of the galaxy.
More recently, Martini et al. (2003) classified Mrk~1210 as a 
tightly wound nuclear dust spiral with the individual
dust lanes traced for over a full rotation about the nucleus.
 
In X-rays, Mrk~1210 is peculiar. It is one of the very few cases of
an AGN caught in a dramatic transition between a Compton-thick, 
reflection dominated state, and a Compton-thin state (Guainazzi et al. 2002). 
This result was possible after comparing XMM Newton observations
with ASCA observations made six years earlier. The transition
is attributed to either a change of the properties of the absorbers
or a ``switching-off'' of the nucleus during the ASCA observations.
In the former scenario, it may indicate a clumpy or patchy structure of
the torus to which the absorbers are associated, while the later may
provide clues about the duty-cycle of the AGNs phenomenon.

From all said above, Mrk~1210 is a key source
where the interplay between an active nucleus and the 
presence of strong star formation can be studied in detail. Therefore, it
is a very important object to investigate the nuclear structure 
of Seyfert~2 galaxies. In order to achieve this goal, it is necessary to
characterize these two components (i.e., nuclear and stellar) 
as completely as possible. The
presence of broad emission line components, a high degree of polarization,
power law near infrared colors, warm far-infrared spectrum, high 
radio brightness temperature in the milli-arcsecond central structure
and, finally, hard X-ray emission, suggest that a significant
contribution comes from the active nucleus. This paper is the first attempt to study
in Mrk~1210 the properties that can be attributed to the AGN,
although not leaving aside the stellar contribution. 
Because of the strong evidence of dust obscuration, the NIR region 
is most appropriate
because it attenuates only a fraction ($\sim 1/7$)
of the light compared to that in the optical, allowing to obtain an 
unbiased estimate of the obscured AGN.

Here, we will study Mrk~1210 by means of JHK spectroscopy
covering the interval 0.8--2.4~$\mu$m, complemented with
optical spectroscopy, aimed at investigating {\it (i)} the 
extinction affecting the nuclear
and extended emitting gas; {\it (ii)} the kinematics of the narrow 
line region (NLR) and {\it (iii)}
the physical properties and conditions of that gas. To meet the first goal,
a variety of extinction diagnostic ratios based on permitted and
forbidden lines, some of them never used before in an AGN, will be used.
This will allow us to make a detailed map the dust distribution on that object.
For the second goal, a detailed comparison of the emission line profiles, 
both in the optical and NIR will be carried out. It will also allow us to
state whether or not a hidden broad line region is in fact present in
the spectra of this object. For the third goal, density and 
temperature sensitive line ratios will be employed to construct a detailed
picture of the ionization structure of emitting gas. In \S ~\ref{data} we 
describe the 
observations and data reduction process. \S ~\ref{features} describes the most
important emission features found in the NIR spectra, some of them reported
for the first time in this work.  \S ~\ref{kinematics} discusses the kinematics
of the NLR gas while \S ~\ref{redd} evaluates and
discusses the dust distribution in the nuclear and extended
emission regions of Mrk~1210; \S ~\ref{physprop} explore the physical condition
throughout the narrow line region and concluding remarks 
are in \S ~\ref{conclusions}. 
A Hubble constant of 75~km~s$^{-1}$~Mpc$^{-1}$ will be used in this work.  


\section{Observations and Data Reduction} \label{data}

\subsection{Infrared Data}
Near-infrared spectra in the interval 0.8--2.4~$\mu$m were obtained 
at the NASA 3~m Infrared Telescope Facility (IRTF) on April 21, 2002 with 
the SpeX facility spectrometer (Rayner et al. 2003). The detector 
consisted of a $1024 \times 1024$ InSb array, with a spatial scale of 
0.15~$\arcsec$/pixel. A $0.8 \arcsec \times 15 \arcsec$ slit was used
during the observations, providing an spectral resolution 
${\rm R} \sim 900$. The slit was oriented at the paralactic angle (60 \degr) 
in order to minimize losses by atmospheric refraction. 
The standard nodding
technique ABBA was employed during the observations. This procedure
allows a good cancellation of sky features and the removal of the bias
and dark levels. After the target, observed with an effective airmass of 1.22, spectra of the AOV star 
HD10961 were taken at similar airmass (1.15) to cancel out the telluric bands and 
flux calibrate the source spectra. Seeing was $0.7 \arcsec$ during the integrations. 
The spectral extraction and calibration procedures were performed using Spextool,
the in-house software developed and provided by the SpeX team for the IRTF 
community (Cushing et al. 2004). When reducing the data, 
each AB source pair was subtracted from each other to remove sky features
and the resulting frames combined to provide a final 2-D spectrum. 
Five 1-D spectra were extracted along the spatial direction by summing up the
signal in a window size of $1 \arcsec$ each. The spectrum associated to the 
nuclear region (hereafter nuclear spectrum, corresponding to the inner
250~pc) was centered in the maximum of the light profile distribution. Two
additional spectra, adjacent to the nuclear one (250~pc SW and NE of the center,
AP1 and AP3, respectively) were extracted at either
side of the light peak distribution. The last two spectra
correspond to regions located at 500~pc SW and NE (AP2 and AP4, respectively)
of the center. Note that at the distance to which Mrk~1210 is located, 
$1 \arcsec =250$~pc.

Telluric absorption correction and flux calibration was applied to
the individual 1-D spectra by means of the IDL based routine {\it xtellcor}
(Vacca et al. 2003). {\it Xtellcor} makes use of the spectrum of the A0V
observed after the source (see above) and a high resolution model of Vega,
to construct a telluric correction spectrum that is free of telluric absorption
features. The flux calibration is done by knowing the magnitude of the 
telluric standard and using the reddening curve of Rieke \& Lebofsky (1985)
in order to estimate the flux as a function of wavelength for the given
star. This response function is then applied to the target.

\begin{figure*}
   \centering
   \includegraphics[width=17cm]{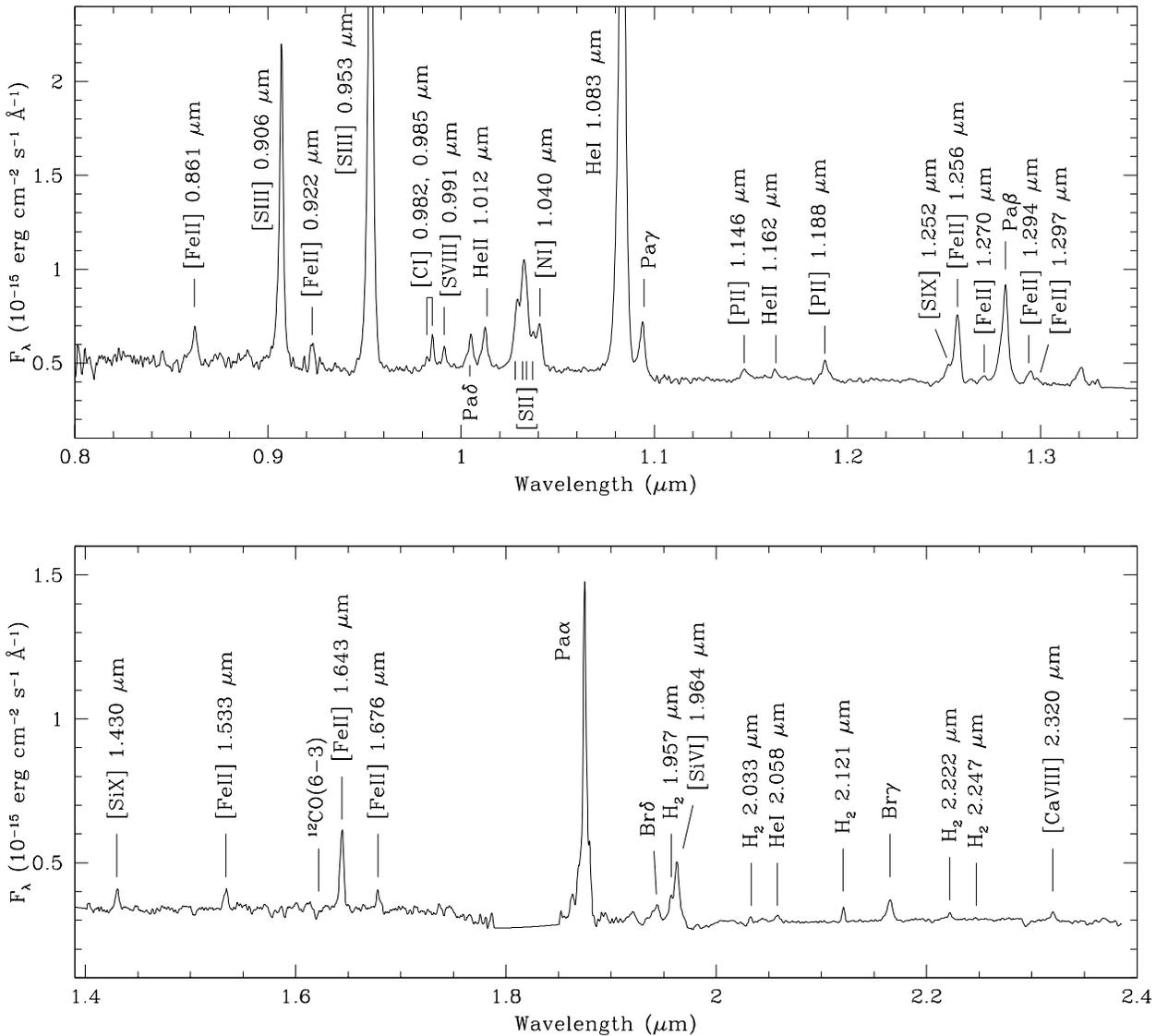}
   \caption{Nuclear spectrum 
of Mrk~1210 in rest wavelength. The labels mark the position of the most important emission lines identified.} 
\label{NUC}
\end{figure*}

\begin{figure*}
 \centering
 \includegraphics[width=17cm]{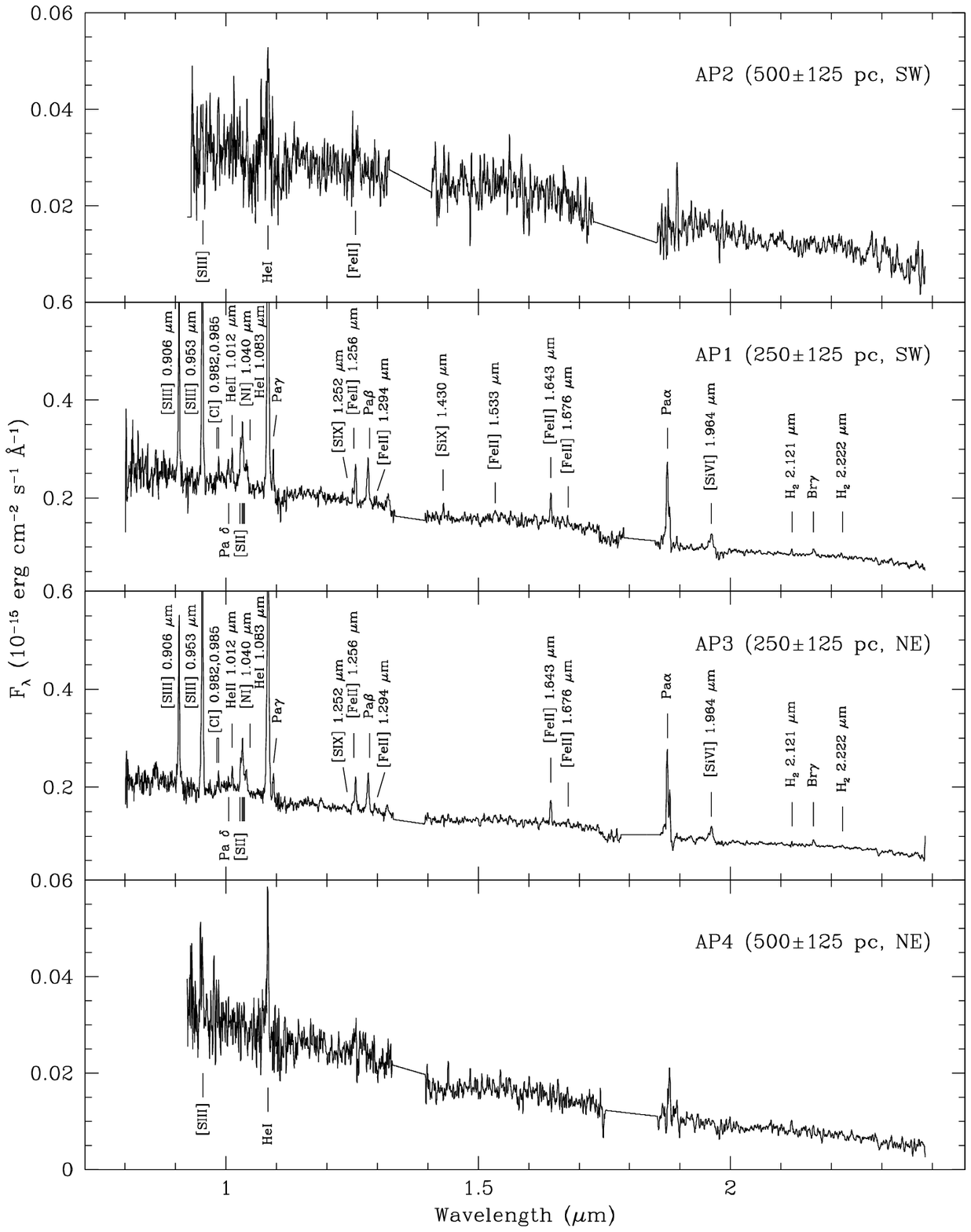}
  \caption{Spectra of the extended emission region of Mrk~1210 in rest wavelength. Each panel represents the spectrum in a rectangular
beam which is $1 \arcsec$ along the slit by the width of the slit ($0.8 \arcsec$). 
The position of the center of each spatial bin, relative to the nucleus, 
is indicated in the upper right of each panel. The second number is the aperture radius.} 
\label{nirspec}
\end{figure*}

Final reduced spectra were corrected for redshift, determined from the average 
$z$ measured from the positions of [\ion{S}{iii}]~0.953~$\mu$m, 
\ion{He}{i}~1.083~$\mu$m, [\ion{Fe}{ii}]~1.256~$\mu$m, 
[\ion{C}{i}]~0.985~$\mu$m, 
Br${\gamma}$ and Pa${\gamma}$. Galactic extinction correction 
of ${\rm E(B-V)}=0.030$ 
was applied to the data (Schlegel et al. 1998). 
The reduced spectra in the interval 0.8--2.4~$\mu$m 
are plotted, in rest wavelengths, in Figures~\ref{NUC} and \ref{nirspec} 
for the nuclear and extended regions, respectively.

\subsection{Optical Data}

\begin{figure*}
   \centering
  \includegraphics[width=17cm]{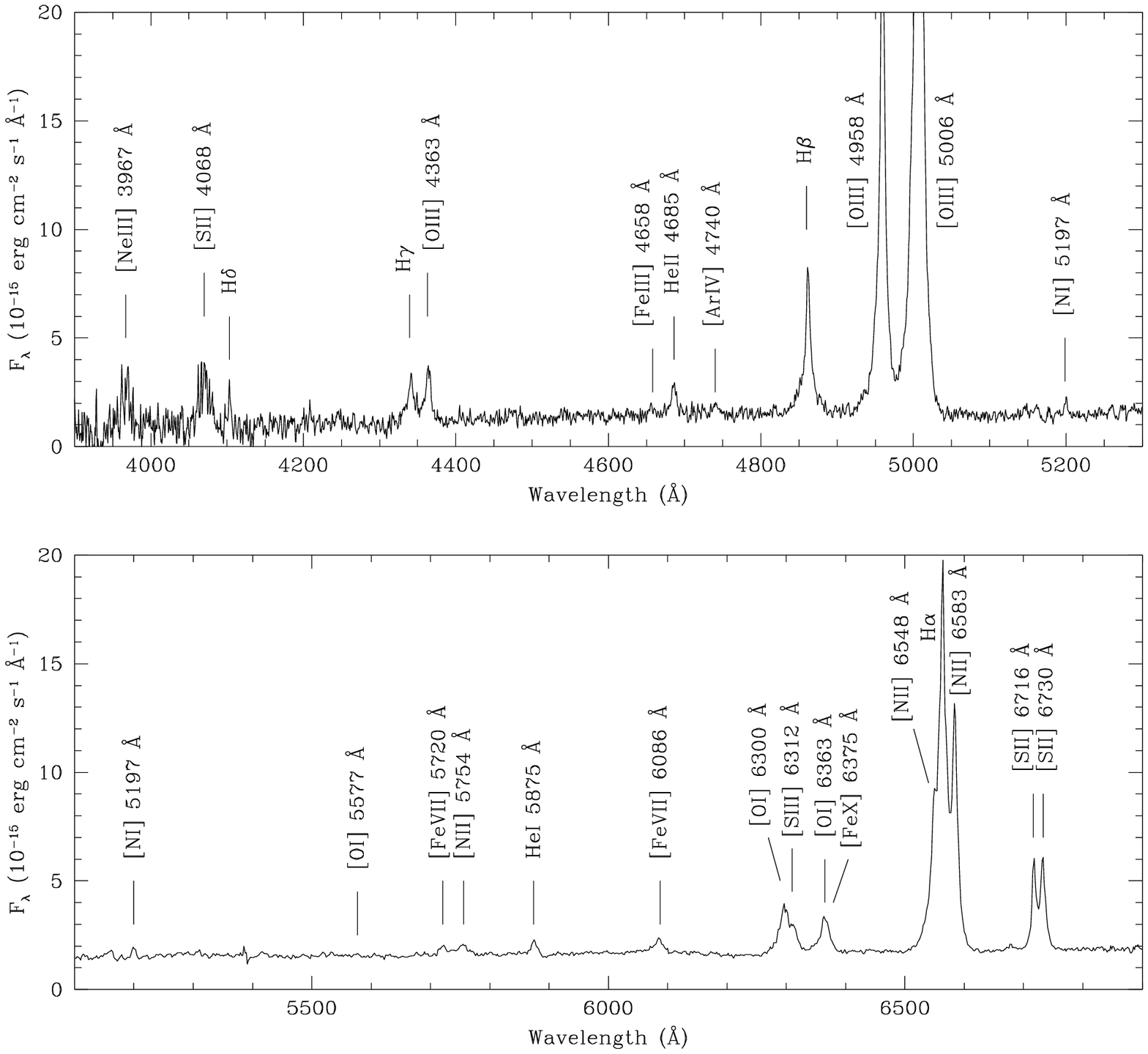}
   \caption{Optical spectrum 
of Mrk~1210 in rest wavelength. The labels mark the position of the most 
important emission lines identified.} \label{optical}
\end{figure*}

In addition to the NIR data, optical spectra covering the wavelength 
interval 3900--6900~\AA\ were taken using two different 
configurations and detectors. The observations were done with the 
Cassegrain spectrograph attached to the 1.6~m
telescope at the Observat\'orio Pico dos Dias, Brazil. 
In both cases, a slit width of $2 \arcsec$, oriented
east-west was employed. The spectrum containing the H$\alpha$ line (5100--6900~\AA,
red spectrum hereafter) was obtained on November 28, 2002 (UT) using
a $1024 \times 1024$ CCD and a 600~l/mm grating, providing a spectral
resolution of 4.5~\AA. The observation covering H$\beta$
(3900--5300~\AA, blue spectrum hereafter), was taken on May 7, 2003
(UT) with a $2048 \times 2048$ CCD
and a 900~l/mm grating. The spectral resolution obtained with the latter
configuration was 2.40~\AA. Data were reduced following IRAF standard
techniques, that is, bias subtraction and flat field division. Wavelength
calibration were done using HeNeAr lamps. For flux calibration,
a spectrophotometric standard was observed after the targets. Due to bad seeing
($\sim 2\arcsec$) the procedure to extract the blue and red spectra was to
integrate all the signal along the slit. 
After reduction and flux calibration, the continuum in the overlap region
(5100--5300~\AA) disagreed. In order to bring both spectra to the same
continuum level, a scale factor of 0.74 was applied to
the blue spectrum. This number corresponds to the ratio of the 
integrated flux of [\ion{N}{i}]~5200~\AA\ for the red and blue spectra.
The choice of [\ion{N}{i}] is because it is the only line common to
both spectra. A Final reduced optical
spectrum is plotted in Figure~\ref{optical}.

Along this manuscript, wavelengths corresponding to the optical data,
from the blue up to 8000~\AA\ will be given in \AA ngstroms. NIR wavelengths 
longwards of to 8001~\AA\ will be reported in microns.

\subsection{Emission line fluxes} \label{elf}
In order to measure the flux of the emission lines in the 
different spectra, we assumed that the line profiles
can be represented by a single or a sum of Gaussian profiles.
The continuum underneath each line was fit by a low-order polynomium,
usually a straight line. 
The LINER routine (Pogge \& Owen 1993) was employed in this process.
Typically, most emission lines were well fitted by a single or a 
sum of two Gaussian components.
Fluxes for the detected lines or upper limits found 
in the nuclear and extended regions in the near-infrared
are listed in Tables~\ref{nirflux} and ~\ref{APflux}.
Table~\ref{optflux} reports the measurements done on the optical spectra. 
In all cases, errors correspond to $3 \sigma$ 
above the continuum level and reflect the uncertainty
in the placement of the continuum and in the S/N
around the line of interest. When a line identification
was uncertain, the Atomic Line List Database\footnote{Available
at http://www.pa.uky.edu/\~\ peter/atomic/} was consulted to 
select a list of candidates. A final identification was
possible according to the presence or absence of additional
lines of the candidate ion. Additional support for line identification
was done by comparing the position of the emission lines with those
observed in planetary nebulae, for instance.

Due to the differences in spatial resolution between the optical and NIR
spectrum, for all the cases were the analysis involved the simultaneous use
of optical and NIR lines, we summed up the fluxes of each line measured 
in the five apertures of the NIR data. 
This procedure was applied for consistency to match the size of 
the aperture used in the extraction of the optical data. This
warrants that the measurements are representative
of the same emitting region.

\begin{table*}
\begin{center}
\caption{Line identifications and fluxes for the infrared spectra.} \label{nirflux}
\begin{tabular}{lcccclcccc} 
\hline\hline
           &      &     & Flux$^*$   &    &&      &     & Flux$^*$   & \\
  \cline{3-5}\cline{8-10}
Ion & $\lambda$~($\mu$m) & AP1 & NUC   & AP3  &Ion & $\lambda$~($\mu$m) & AP1 & NUC   & AP3 \\
  \hline\hline
$\lbrack$\ion{Fe}{ii}$\rbrack$ & 0.861         &--& $4.7 \pm 0.1$ & -- &
$\lbrack$\ion{Fe}{ii}$\rbrack$ & 1.256         & $2.7 \pm 0.2$ & $11.2 \pm 0.1$  & $2.2 \pm 0.1$ \\

$\lbrack$\ion{S}{iii}$\rbrack$ & 0.906         & $11.0 \pm 0.4$ & $50.0 \pm 1.5$ & $7.6 \pm 0.5$ &
$\lbrack$\ion{Fe}{ii}$\rbrack$ & 1.270         &--& $1.0 \pm 0.2$ & --\\

$\lbrack$\ion{Fe}{ii}$\rbrack$ & 0.923         &--& $3.8 \pm 0.3$ & --&
Pa$\beta$                      & 1.281         & $4.8 \pm 0.3$ & $23.6 \pm 0.3$ & $3.6 \pm 0.4$ \\

$\lbrack$\ion{Fe}{ii}$\rbrack$ & 0.946         &--& $1.6 \pm 0.5$ & --&
$\lbrack$\ion{Fe}{ii}$\rbrack$ & 1.294         & $0.2 \pm 0.1$ & $2.2 \pm 0.1$ & $0.5 \pm 0.2$ \\

$\lbrack$\ion{S}{iii}$\rbrack$ & 0.953         & $25.6 \pm 0.5$ & $120.1 \pm 0.8$  & $22.7 \pm 0.5$ &
$\lbrack$\ion{Fe}{ii}$\rbrack$ & 1.297         & $0.3 \pm 0.1$ & $0.7 \pm 0.1$ & $0.2 \pm 0.1$ \\

$\lbrack$\ion{C}{i}$\rbrack$   & 0.982         & $0.5 \pm 0.3$ & $1.3 \pm 0.2$    & $0.3 \pm 0.2$ &
$\lbrack$\ion{Si}{x}$\rbrack$  & 1.430         & $0.7 \pm 0.2$ & $2.8 \pm 0.4$ & -- \\

$\lbrack$\ion{C}{i}$\rbrack$   & 0.985         & $1.0 \pm 0.3$ & $3.3 \pm 0.2$    & $0.8 \pm 0.2$ &
$\lbrack$\ion{Fe}{ii}$\rbrack$ & 1.533         & $0.6 \pm 0.3$ & $2.8 \pm 0.1$ & -- \\

$\lbrack$\ion{S}{viii}$\rbrack$& 0.991         &--& $2.9 \pm 0.2$ & --&
$\lbrack$\ion{Fe}{ii}$\rbrack$ & 1.643         & $1.9 \pm 0.2$ & $12.5 \pm 0.5$ & $1.7 \pm 0.2$ \\

Pa$\delta$                     & 1.004         & $1.6 \pm 0.4$ & $5.3 \pm 0.5$ & $1.0 \pm 0.3$ &
$\lbrack$\ion{Fe}{ii}$\rbrack$ & 1.676         & $0.5 \pm 0.1$ & $2.5 \pm 0.3$ & $0.5 \pm 0.1$ \\

\ion{He}{ii}                   & 1.012         & $2.4 \pm 0.4$ & $7.1 \pm 0.3$ & $1.0 \pm 0.3$ &
Pa$\alpha$                     & 1.875         &&&\\

$\lbrack$\ion{S}{ii}$\rbrack$  & 1.028         & $2.8 \pm 0.3$ & $11.3 \pm 0.9$ & $1.8 \pm 0.2$ &
Br$\delta$                     & 1.944         & $0.5 \pm 0.3$ & $3.8 \pm 1.1$ & --\\

$\lbrack$\ion{S}{ii}$\rbrack$  & 1.032         & $2.8 \pm 0.3$ & $12.2 \pm 0.9$   & $2.7 \pm 0.2$ &
H$_2$                          & 1.957         & $0.7 \pm 0.3$ & $2.5 \pm 0.3$ & $0.5 \pm 0.1$ \\

$\lbrack$\ion{S}{ii}$\rbrack$  & 1.033         & $2.3 \pm 0.3$ & $12.2 \pm 0.9$   & $0.9 \pm 0.2$ &
$\lbrack$\ion{Si}{vi}$\rbrack$ & 1.964         & $1.9 \pm 0.4$ & $13.3 \pm 0.9$ & $1.6 \pm 0.2$ \\

$\lbrack$\ion{S}{ii}$\rbrack$  & 1.037         & $1.0 \pm 0.3$ & $5.6 \pm 0.9$   & $0.8 \pm 0.2$ &
H$_2$                          & 2.033         &--& $0.5 \pm 0.1$ &--\\

$\lbrack$\ion{N}{i}$\rbrack$   & 1.040         & $2.3 \pm 0.3$ & $8.7 \pm 1.0$    & $1.1 \pm 0.2$ &
\ion{He}{i}                    & 2.058         &--& $1.3 \pm 0.3$ &--\\

\ion{He}{i}                    & 1.083         & $37.8 \pm 0.5$ & $179.3 \pm 0.5$  & $30.0 \pm 0.4$ &
H$_2$                          & 2.121         & $0.2 \pm 0.1$ & $1.6 \pm 0.8$ & $0.10 \pm 0.03$ \\

Pa$\gamma$                     & 1.093         & $2.5 \pm 0.4$ & $10.5 \pm 0.5$  & $1.9 \pm 0.5$ &
Br$\gamma$                     & 2.165         & $0.8 \pm 0.1$ & $4.6 \pm 0.2$ & $0.6 \pm 0.3$ \\

$\lbrack$\ion{P}{ii}$\rbrack$  & 1.146         &--& $2.3 \pm 0.3$ & -- &
H$_2$                          & 2.223         & $0.3 \pm 0.1$ & $0.9 \pm 0.1$ & $0.15 \pm 0.07$ \\

H$_2$                          & 1.185         &--& $0.25 \pm 0.14$ & --&
H$_2$                          & 2.247         &--& $0.2 \pm 0.1$ & --\\

$\lbrack$\ion{P}{ii}$\rbrack$  & 1.188         &--& $3.7 \pm 0.3$ & --&
$\lbrack$\ion{Ca}{viii}$\rbrack$& 2.320        &--& $1.6 \pm 0.1$ & -- \\

$\lbrack$\ion{S}{ix}$\rbrack$  & 1.252         & $1.0 \pm 0.2$ & $4.2 \pm 0.5$ & $0.5 \pm 0.2$ 
&&&&\\
\hline
\hline
   \end{tabular}
\end{center}
$^*$ In units of 10$^{-15}$ erg cm$^{-2}$ s$^{-1}$.
\end{table*}

\begin{table}
\begin{center}
\caption{Line identifications and fluxes, in the NIR, for the AP2 and AP4.}\label{APflux}
\begin{tabular}{lccc}
  \hline\hline
 & & \multicolumn{2}{c}{Flux$^*$}  \\
  \cline{3-4}
Ion & $\lambda$ ($\mu$m) & AP2 & AP4 \\
  \hline\hline

$\lbrack$\ion{S}{iii}$\rbrack$ & 0.953 & $< 0.53$  & $0.60 \pm 0.3$ \\
\ion{He}{i}                    & 1.083 & $1.08 \pm 0.4$  & $0.97 \pm 0.15$ \\
$\lbrack$\ion{Fe}{ii}$\rbrack$ & 1.256 & $< 0.63$       & $< 0.2$  \\
\hline
\hline
\end{tabular}
\end{center}
$^*$ In units of 10$^{-15}$ erg cm$^{-2}$ s$^{-1}$.
\end{table}

\begin{table}
\begin{center}
\caption{Line identifications and fluxes for the  optical spectra.}\label{optflux}
\begin{tabular}{lcclcc} 
  \hline\hline
Ion & $\lambda$~(\AA) & Flux$^*$ & Ion & $\lambda$~(\AA) & Flux$^*$ \\
  \hline\hline
$\lbrack$\ion{Ne}{iii}$\rbrack$& 3967 & $ 22.8 \pm 9.0$ &
$\lbrack$\ion{N}{ii}$\rbrack$   & 5754  & $6.3 \pm 1.0$ \\

$\lbrack$\ion{S}{ii}$\rbrack$   & 4068  & $ 36.4 \pm 6.8$ &
\ion{He}{i}                     & 5875  & $8.0 \pm 1.0$ \\ 

H$\delta$                       & 4101 & $ 9.3 \pm 6.2$ &
$\lbrack$\ion{Fe}{vii}$\rbrack$ & 6086  & $9.5 \pm 1.2$ \\

H$\gamma$                       & 4340 & $ 22.2 \pm 3.3$ &
$\lbrack$\ion{O}{i}$\rbrack$    & 6300  & $45.8 \pm 1.5$  \\

$\lbrack$\ion{O}{iii}$\rbrack$  & 4363  & $ 22.2 \pm 4.0$ &
$\lbrack$\ion{S}{iii}$\rbrack$  & 6312  & $13.2 \pm 1.0$  \\

\ion{He}{ii}                    & 4685  & ${ \b 10.3 \pm 2.6}$ &
$\lbrack$\ion{O}{i}$\rbrack$    & 6363  & $33.0 \pm 1.5$ \\

H$\beta$                        & 4861  & $ 72.6 \pm 3.5$ &
$\lbrack$\ion{N}{ii}$\rbrack$   & 6548  & $19.0 \pm 1.3$ \\

$\lbrack$\ion{O}{iii}$\rbrack$  & 4958  & $ 221.8 \pm 2.2$ &
H$\alpha$                       & 6562  & $419.3 \pm 1.8$ \\

$\lbrack$\ion{O}{iii}$\rbrack$  & 5006  & $ 636.4 \pm 2.2$ &
$\lbrack$\ion{N}{ii}$\rbrack$   & 6583  & $57.0 \pm 1.3$  \\

$\lbrack$\ion{N}{i}$\rbrack$    & 5197  & $4.0 \pm 0.6$ &
$\lbrack$\ion{S}{ii}$\rbrack$   & 6716  & $38.4 \pm 1.1$ \\

$\lbrack$\ion{O}{i}$\rbrack$    & 5577  & $< 1.8$  &
$\lbrack$\ion{S}{ii}$\rbrack$   & 6730  & $42.4 \pm 1.1$  \\

$\lbrack$\ion{Fe}{vii}$\rbrack$ & 5720  & $8.3 \pm 2.0$ \\
\hline
\hline
   \end{tabular}
 \end{center}
$^*$ In units of 10$^{-15}$ erg cm$^{-2}$ s$^{-1}$.
\end{table}


\section{The near-infrared spectrum of Mrk~1210} \label{features}

The nuclear spectrum of Mrk~1210 displays a plethora
of emission lines, from those emitted from molecular H$_{2}$
and low and medium ionization species such as [\ion{C}{i}],
[\ion{S}{ii}] and [\ion{S}{iii}], up to very high ionization lines of 
[\ion{S}{ix}] and [\ion{Si}{x}]. This implies a very large range of physical 
conditions of the emitting gas, even within the central 250~pc, to allow the
simultaneous coexistence of molecules, somehow protected from 
the intense radiation
of the central engine, and highly ionized gas, undoubtedly associated to
the AGN phenomenon. In addition, absorption lines of
\ion{Ca}{ii} and CO are also observed, indicating the presence of a 
circumnuclear stellar population. 

The off-nuclear spectra of Mrk~1210 are also peculiar. At 250~pc NE
and SW of the nucleus, emission from \ion{H}{i} and \ion{He}{i} dominates.
Farther out, at 500~pc from the center, both the NE and SW spectra
show emission from \ion{He}{i}, [\ion{Fe}{ii}] and [\ion{S}{iii}]. Weak 
Pa$\beta$ is detected in the 500~pc NE, suggesting that the extended 
emission is dominated by matter-bounded clouds. The 
detection of extended [\ion{S}{iii}] is compatible with the data 
presented by Fraquelli et al. (2003) who show extended [\ion{O}{iii}]~5006~\AA\
up to 1~kpc from the nucleus.

\subsection{The forbidden NIR spectrum}

The brightest forbidden lines observed in the nuclear spectrum of Mrk~1210 are
[\ion{S}{iii}]~0.953~\micron\ and [\ion{S}{iii}]~0.906~\micron. In
addition, Figure~\ref{nirspec} shows that the former line is not limited to the
nuclear region. It extends up to 500~pc NE and 250~pc SW from the center. 
In the AP2 spectrum, a 
feature at the expected position of [\ion{S}{iii}]~0.953~\micron\
is detected but the S/N is low. For this reason, we only report 
an upper limit for its flux.

An inspection of Figure~\ref{NUC} shows that Mrk~1210 displays a very 
rich coronal NIR spectrum. Here, we report the first detection of
[\ion{S}{viii}]~0.991~$\mu$m, 
[\ion{S}{ix}]~1.252~$\mu$m, 
[\ion{Si}{x}]~1.430~$\mu$m, and 
[\ion{Ca}{viii}]~2.320~$\mu$m on this object. The strongest coronal line 
observed is [\ion{Si}{vi}] 
which, to our knowledge, is the only one   
previously reported for this galaxy in the 0.8--2.4~$\mu$m range 
(Rodr\'{\i}guez-Ardila et al. 2004). 
At 250~pc from the nucleus, high-ionization lines such 
as [\ion{S}{vi}]~1.964~$\mu$m and 
[\ion{Si}{x}]~1.430~$\mu$m are still visible, mostly in the SW direction. This,
however, cannot be taken as definitive probe of extended emission in the
coronal lines. It is possible that they represent scattered nuclear photons 
or even that they are emitted in the inner few tens of parsecs of the 
extended region.

In addition to the 
features mentioned above, Mrk~1210 is characterized by a conspicuous 
[\ion{Fe}{ii}] spectrum, not previously reported in the literature 
for other AGN. Besides 
[\ion{Fe}{ii}]~1.256~$\mu$m and 1.643~$\mu$m, common in most Seyferts 
(Rodr\'{\i}guez-Ardila et al. 2004), and whose flux ratio is an 
excellent reddening indicator for the NLR, up to eight more lines are detected
in the J and H-bands. The presence of these lines 
is a prime opportunity to use the forbidden iron lines as
diagnostic tool of the region where they are formed.  
In Section~\ref{redd} we will 
explore the diagnostic capabilities of the [\ion{Fe}{ii}] as a probe of
the physical conditions of the NLR. 
As already mentioned, we found that [\ion{Fe}{ii}]~1.256~$\mu$m
extends up to 500~pc NE and SW from
the center (see Figure~\ref{nirspec}). 

Other interesting features on the NIR forbidden spectrum 
are [\ion{S}{ii}]~1.028~$\mu$m, 1.032~$\mu$m, 1.033~$\mu$m
and 1.037~$\mu$m\footnote{We will use the term [\ion{S}{ii}]~1.03~$\mu$m, 
hereafter, to indicate the sum of the fluxes of these four 
sulfur lines}. These lines, along 
with the optical doublet [\ion{S}{ii}]~4068,~4076~\AA\, arise from 
the same upper term. It means that
the flux ratio between the NIR and optical sulfur lines depends 
primarily on their 
transition probabilities. Therefore, it can be used as an additional
reddening indicator.

Finally, low ionization lines of [\ion{N}{i}], [\ion{C}{i}] and 
[\ion{P}{ii}] are also detected in the nuclear spectrum. With 
exception of the later, these lines are also observed in AP1 and 
AP3 (see Fig.~\ref{nirspec}). In 
the nucleus, the flux ratio
[\ion{P}{ii}]/[\ion{C}{i}] is 1.3 (see Table~\ref{nirflux}).
The detection of [\ion{P}{ii}] lines, of considerable strength, 
is surprising. This is because in the sun, phosphorus is a factor 
$\sim1\,000$ times less abundant than carbon. 
Therefore, if the P/C abundance is even closer to solar, the 
[\ion{P}{ii}] lines should not be present unless carbon and other
strong abundant elements are much more optically thick than
they appear. An alternative to this apparent contradiction is
to assume that phosphorus is over-abundant
by a factor 10-30 compared to its solar abundance, relative
to hydrogen. A similar problem is found in high-redshift
quasars, where broad absorption lines of 
\ion{P}{v}~$\lambda\lambda$1118,1128~\AA\ are detected 
(Hamman 1998, Hamman et al. 2002). The NIR [\ion{P}{ii}] lines
may probably help to set some constraints to the abundance
of phosphorus in AGN. This subject, however, is out of the
scope of this paper.

\subsection{Molecular and absorption lines}
Several molecular hydrogen lines are present in the nuclear  
and extended NIR spectra of Mrk~1210, mostly in the K-band. 
We detected H$_2$ emission up to a distance of 250~pc, both
NE and SW from the center. An inspection to the spectra of
Figs.~\ref{NUC} and~\ref{nirspec} shows that the strongest 
H$_2$ lines observed in this galaxy are (1,0)S(1) at 2.121~$\mu$m
and (1,0)S(3) at 1.957~$\mu$m. Also detected in the K-band but less
intense are (1,0)S(2)~2.033~$\mu$m, (1,0)S(1)~2.223~$\mu$m
and (2,1)S(1)~2.247~$\mu$m.
These lines were studied by Rodr\'{\i}guez-Ardila et al. (2004)
in a sample of AGN, including Mrk~1210, aimed at determining 
the dominant H$_2$ excitation mechanisms.    
They found that, for Mrk~1210, the H$_2$ is predominantly 
excited by stellar processes (UV heating), with some contribution of
fluorescence ($\sim 30$\%) and little influence of X-ray heating
(as is the dominant mechanism for most AGN). 
This result is supported by the presence of a circumnuclear starburst 
(Schulz \& Henkel 2003) and strong H$_2$O megamaser emission 
coming from a small region around the nucleus (Braatz \& 
Wilson 1994) that can be associated with star forming regions.
The vibrational and rotational temperatures derived by 
Rodr\'{\i}guez-Ardila et al. (2004) for the nuclear
H$_2$ gas were of $3100 \pm 700$~K and $900 \pm 300$~K, respectively.  

The flux of the H$_2$ (1,0)S(1)~2.121~$\mu$m, listed in 
Table~\ref{nirflux} for the different apertures and corrected for 
reddening (see Sec.~\ref{redd}), can be used to derive the mass 
of hot molecular gas present in the inner few hundred parsecs 
of Mrk~1210. To this purpose, we used the expression:

\begin{center}
$ m_{\rm H_{2}} \simeq 5.0875 \times 10^{13} D^{2}I_{1-0S(1)} $ \\
\end{center}

\hspace{-0.60cm} taken from Reunanen et al. (2002) and assuming ${\rm T}=2\,000$~K,
a transition probability $A_{S(1)}=3.47 \times 10^{-7}$~s$^{-1}$
(Turner et al. 1977) and the population fraction in the $\nu =1$, $J =3$
level $f_{\nu=1, J=3}=0.0122$ (Scoville et al. 1982). $I_{1-0S(1)}$ is the 
flux of \h2~2.121~$\mu$m corrected for intrinsic ${\rm E(B-V)}$. 

The values of the \h2\ mass found are $260 \pm 40$~M$\odot$ for the nucleus,
$30 \pm 13$~M$\odot$ for AP1 and $14 \pm 4$~M$\odot$ AP3, totalizing
nearly 300~M$\odot$ of hot molecular gas for the inner 500~pc of
Mrk~1210. It is in very good agreement with the nuclear \h2\ 
masses reported by Reunanen et al. (2003) for a sample of Seyfert~1 
and 2 galaxies.
The small mass of hot \h2\ contrasts with the cold molecular mass of
$7.94 \times 10^{8}$~M$\odot$ found by Raluy et al. 
(1998) from CO($1\rightarrow 0$) observations and using the standard
conversion factor found in Galactic giant molecular clouds,
$M_{\rm H_{2}} = 5.8$~$L_{\rm CO}$. The aperture of the CO data 
covers the inner 5.7~kpc. Interestingly, Raluy et al. (1998) found a
high star formation efficiency in Mrk~1210 (derived from the 
$L_{\rm IR}/M_{\rm H_{2}}$ ratio), comparable to that of
normal galaxies like the Milky Way or M~33. In fact, from a list
of 12 AGNs, Mrk~1210 is the one that shows the second largest 
star formation efficiency. They however, 
called the attention to the fact that a significant fraction 
of the IR luminosity (up to 50\%) could come from the AGN
itself, resulting in an overestimation of the star formation
efficiency by up to a factor of 2. \\

A rough estimate of the contribution of the stellar population
flux to the H-band can be made by measuring the depth of
the $^{12}{\rm CO(6-3)}$ overtone bandhead at 1.618~$\mu$m (see Fig.~\ref{NUC}). 
We found that the observed
line depth is $11\% \pm 1\%$ of the continuum. This can be
compared to the $\sim 20\%$ that is tipically expected for a
population of GKM supergiants that dominate the H-band
light for stellar populations older that $10^{7}$~yr 
(Schinnerer et al. 1998). It means that $55\% \pm 5\%$ of the total 
flux in the H-band continuum is due to GMK supergiants.
Following Schinnerer et al. (1998), although these stars dominate
the stellar emission in the H-band, about one-third of the
total stellar flux is contributed by stars of other stellar classes.
Therefore, $83\% \pm 8\%$  of the H-band continuum
in a $0.8\arcsec \times 1\arcsec$ aperture is of stellar origin.
This result is in very good agreement to the $71\% \pm 6\%$ contribution
of stellar population older that $10^{7}$~yr determined by 
Cid Fernandes et al. (2001) for this object, based 
on optical spectroscopy. It means that the H-band continuum is essentially dominated 
by the stellar population, with only $\sim$20\% due to the AGN. This new value
refines the claim made by Raluy et al. (1998), who estimated
that at least 50\% of the IR could come from the active
nucleus. The above calculations suggest that NIR spectroscopy
is a potential tool to unveil hidden starburst components, either
in highly obscured objects or in those in which the optical 
continuum emission is dominated by the AGN component (as in Seyfert 1
galaxies).

Even after correcting for the increased contribution of the
stellar population to the continuum, Mrk~1210 continues 
to display one of the largest (if not the largest) star formation 
rates among the group of 12 AGN studied by Raluy et al. (1998).

The strong contribution of stellar light to the Mrk~1210 continuum
in the inner $0.8 \arcsec^{2}$ is also compatible with the analysis of
the \h2\ lines presented in the beginning of this section. It
further supports the hypothesis 
that in this AGN the stellar population contributes significantly
to the excitation of the molecular lines.


\section{Kinematics of the narrow line region} \label{kinematics}

\begin{figure*}
   \centering
  \includegraphics [width=12cm]{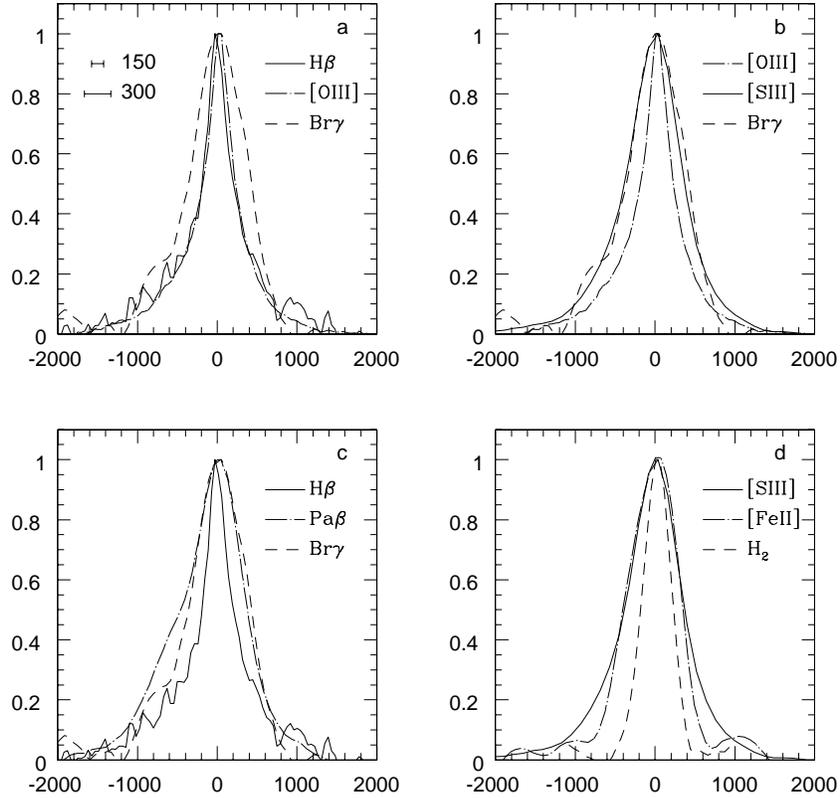}
   \caption {Profiles of nuclear lines in Mrk~1210 in velocity space. 
The lines are normalized to the same peak flux. The small bars in panel
(a) represent the instrumental profile of the blue spectrum (FWHM=150~\kms)
and that of the NIR region (FWHM=330~\kms).}\label{profile1}
\end{figure*}

The large amount of spectral information at medium resolution, 
covering the optical and NIR regions, allows us to study the
form and width of the emission lines in Mrk~1210. The
purpose is to map the velocity field of the NLR, examine the
presence of genuine BLR components and estimate the
most probable location of the emitting gas. 

The results of the
Gaussian decomposition carried out to measure the line fluxes (see 
Sec.~\ref{elf}) showed that most lines were well-fitted by 
a single narrow Gaussian component of 
${\rm FHWM} \sim 500$~km~s$^{-1}$. However,
bright lines such as [\ion{S}{iii}], [\ion{O}{iii}], Pa$\beta$, 
and \ion{He}{i} needed
an additional, broader blue-shifted component (see Table~\ref{widths}),
of ${\rm FWHM} > 1\,000$~km~s$^{-1}$. It is known that at 
typical BLR electron densities ($n_{\mathrm e} \sim 10^{9-11}$~cm$^{-3}$) only 
broad permitted lines (with ${\rm FWHM} \sim 10^{3-4}$~km~s$^{-1}$) 
would be present. In this galaxy not only
the permitted lines have broad components but also the forbidden ones.
Veilleux et al. (1997) reported that the Pa$\beta$ and Br$\gamma$ 
lines profiles of Mrk~1210 were characterized 
by a strong narrow component on top of a broad base. They compared
the \ion{H}{i} lines with
[\ion{Fe}{ii}]~1.256~$\mu$m and H$_2$~2.121~$\mu$m. Because of
the smaller FWHM displayed by the latter two lines, they
suggested that some of the emission was
produced in a genuine high density BLR. With our data we are able to extend 
this comparison to a larger number of permitted and forbidden lines to
see if the hidden broad line region proposed by Veilleux et al. (1997)
is in fact observable in the NIR spectrum of Mrk~1210.

Figure~\ref{profile1} shows a comparison of optical and 
NIR emission line profiles, in velocity space. The upper left panel (a) 
shows [\ion{O}{iii}]~5006~\AA, H$\beta$ and Br$\gamma$
while the upper right panel (b) plots [\ion{S}{iii}]~0.953~\micron, 
[\ion{O}{iii}]~5006~\AA, and Br$\gamma$. It can be seen that
both permitted and forbidden lines in the NIR region are 
quite similar in form and width, ruling out the hypothesis of a 
broad component associated to the BLR. Moreover, 
Figure~\ref{profile1} shows that the optical lines are systematically 
narrower at FWHM than the NIR ones. In contrast, the full width at 
zero intensity (FWZI) are rather similar in both regions. In
panel (c) a comparison of H$\beta$, Pa$\beta$ and Br$\gamma$ is shown.
Note, again, that the FHWM of the optical and NIR lines are different
but the FWZI is equivalent. 

The difference in FWHM among the narrow components may be 
attributed to either selective extinction or to 
a difference in the instrumental resolution. In the first scenario, 
optical photons emitted by low-velocity clouds 
are more absorbed than the NIR ones, allowing 
us to see preferentially the optical emission originated 
in high velocity clouds, which apparently are less subjected
to extinction. In the second scenario, the difference
in FWHM between NIR and optical lines is due to
differences in the instrumental profiles. In fact, the instrumental
FWHM of the NIR data is a factor of about 2.2 larger than that of the blue
spectrum (instrumental FWHM = 150 \kms). Although at first sight
the latter argument would be sufficient to explain the differences in
width, the values of FHWM listed in Table~\ref{widths} point out that very
likely selective extinction dominates. This is because in 
Table~\ref{widths}, the FWHM of the lines have been corrected
for instrumental resolution. Clearly, all the lines are spectroscopically
resolved, and most importantly, the narrow components of 
[\ion{O}{iii}] and H$\beta$ are narrower than those of
[\ion{S}{iii}] and Pa$\beta$, for instance.     

Note that the Pa$\beta$ profile shows enhanced emission   
blueward of the line center. This feature is
visible not only in Mrk~1210, but also in other Seyfert galaxies 
(e.g. NGC~2110, Storchi-Bergman et al. 1999; NGC~4151, Thompson 1995). We
attribute this asymmetry primarily to the presence of
[\ion{Fe}{ii}]~1.278~$\mu$m, which contaminates the blue wing
of Pa$\beta$. In spite of this feature, the Pa$\beta$ and Br$\gamma$ 
line profiles are rather similar, and the difference at FWHM with 
the optical lines is clear. In panel (d) we compare 
[\ion{Fe}{ii}]~1.643~$\mu$m, H$_2$~2.121~$\mu$m and 
[\ion{S}{iii}]~0.953~\micron. Note the slight blue 
asymmetry in [\ion{S}{iii}], not present neither in 
[\ion{Fe}{ii}] nor in H$_2$. It can also be seen that from 
peak intensity to FWHM, [\ion{Fe}{ii}] and [\ion{S}{iii}] are 
indistinguishable. From FWHM to FWZI, the two profiles diverge, 
with [\ion{S}{iii}] becoming broader. In contrast, H$_2$ is 
significantly narrower, almost half the FWHM value of the former 
two. Since extinction cannot be invoked to explain these 
differences because all three lines are located in the NIR region, 
the most plausible explanation is that the bulk of [\ion{S}{iii}], 
[\ion{Fe}{ii}] and H$_2$ originates from different volumes of gas, 
subjected to different dispersion velocity conditions.

The results of the above comparison show that the same broad component 
present in the permitted lines is also detected in 
some of the forbidden lines. This indicate that the gas responsible
for the broad lines is most likely associated to the NLR itself. 
Additional support to this picture is found from the work 
of Lutz et al. (2002). They compared the emission line profiles
of [\ion{Si}{ix}]~3.934~$\mu$m 
and Br$\alpha$ in Mrk~1210 and decided against the BLR interpretation 
because the two profiles were 
indistinguishable within the S/N limitations.

We propose that the broad component seen in [\ion{S}{iii}],
[\ion{O}{iii}] and the permitted lines is associated
to outflows, instead of the hidden BLR suggested by Veilleux.
Evidence of this scenario comes from Middelberg et al. (2004), who carried
out radio observations of Mrk~1210 at 18~cm and 6~cm with 
EVN, MERLIN and VLBA. The EVN and VLBA 18~cm images reveal a
bright compact object and a weaker component toward the south-east
at a distance of 8.6~pc. They marginally detect a third component
at a distance of 30.6~pc from the bright compact source. They suggest that
the later component could probably be the continuation of the radio
ejecta. Moreover, their 6~cm VLBA data resolves the brightest source
into an arc of four components as well as the south-east feature. 
Note that most of the galaxy's 6~cm emission comes from scales
between 3~pc and 16~pc. This region is unresolved from our data.
Middelberg et al. (2004) suggest three scenarios for the
radio structure in Mrk~1210. A free-free emitting disk, synchrotron
emission from a torus or extended accretion disk and a system
composed of a core and an outflow. Based on physical arguments,
the latter hypothesis is most favored. The resolved arc of
four components could be bright knots due to a shock in an
extended, low brightness temperature radio outflow, with 
similar conditions to those in NGC~5506, NGC~1068 and Mrk~231.  
 
An additional support to this picture is that the peak of
the broad component detected in [\ion{S}{iii}], [\ion{O}{iii}], \ion{H}{i}
and \ion{He}{i} is blueshifted relative to the systemic velocity
of the galaxy (see Table~\ref{widths}). The shift
in the forbidden lines is similar to that of \ion{He}{i} and
almost half of \ion{H}{i}. This broad component is probably associated to
the series of four knots reported by Middelberg et al. (2004), not
spatially resolved in our nuclear spectrum. This
also would explain the fact that the line profiles of  some
forbidden and permitted lines in Figure~\ref{profile1} are different
at FWHM but similar at FWZI. The largest dispersion velocities
are associated to the outflow component, with little or no extinction
toward the line of sight. At FWHM, the bulk of the line profile
is dominated by the emission from the classical NLR clouds. The
NIR lines are broader because we are looking deep into the
clouds, where the dispersion velocity is larger.

\begin{table}
\caption{FWHM$^{*}$ and shifts of the peak centroid of the broad component measured in the lines displaying narrow and broad components.} \label{widths}

       \begin{center}

  \begin{tabular}[t]{lccccc}
  \hline\hline
 & & \multicolumn{2}{c}{FWHM [km~s$^{-1}$]}  &   \\
\cline{3-4}
 Ion & $\lambda$~($\mu$m) & BC$^1$ & NC$^2$ & $\Delta$V   \\
  \hline\hline
H$\beta$ & 0.486 & 1410 & 330 & 234 \\
$\lbrack$\ion{O}{iii}$\rbrack$ & 0.495 & 1130 & 260 & 76\\
$\lbrack$\ion{O}{iii}$\rbrack$ & 0.500 & 1120 & 260 & 78\\
$\lbrack$\ion{S}{iii}$\rbrack$ & 0.906 & 1260 & 460 & 114\\
$\lbrack$\ion{S}{iii}$\rbrack$ & 0.953 & 1270 & 460 & 102\\
\ion{He}{i} & 1.083 & 1510 & 500 & 113 \\
Pa$\beta$ & 1.281 & 1360 & 390 & 225 \\
\hline
\hline
\end{tabular}
\end{center}
$^*$ Corrected for instrumental resolution.\\
$^1$ Broad Component\\
$^2$ Narrow Component
\end{table}


\section{The extinction affecting the NLR of Mrk~1210} \label{redd}

The numerous \ion{H}{i} and forbidden lines in the
spectra of Mrk~1210, spanning a large interval in wavelength,
allowed us to evaluate the intrinsic extinction
affecting the NLR within the inner few hundred parsec 
by means of several indicators. 
In the cases were the determination of reddening involved only NIR lines  
we are able to trace it at different distances from the nucleus, 
obtaining additional information about the dust distribution.

The extinction was determined from the comparison of the predicted and 
observed emission line ratios, assuming the standard extinction law of 
Cardelli et al. (1989) for ${\rm R_V}=3.1$ (note, however, 
that at these wavelengths,  
the extinction law is nearly independent of the assumed R$_{\rm V}$). 
Listed in Table~\ref{redd} are the line ratios 
used for the reddening determinations (first column), the intrinsic 
ratio values (second column), and the measured line ratios and estimated 
reddening for three infrared spectra (AP1, NUC and AP3). For AP2 and AP4,
no sufficient information is available. The last two 
rows in Table~\ref{redd} list the ratios involving optical lines. These latter 
values correspond to an emitting region size equivalent to the one 
covered by the sum of the five infrared 
extractions (500~pc). In all cases, the errors correspond to $3\sigma$ and 
were determined assuming that the only quantities that introduce uncertainties 
are the measured line fluxes.

For the hydrogen lines, we used the intrinsic ratios 
for case B given by Osterbrock (1989).
Our results point out to extinction variations up to a factor 
of nearly 5 within the NLR. 
Recall, however, that the ${\rm E(B-V)}$ derived
using Pa$\beta$ may be misleading. This is because of  
the contamination 
of the Pa$\beta$ line with [\ion{Fe}{ii}]~1.278~$\mu$m,
mentioned in the previous section. 
This feature increases the Pa$\beta$ flux, leading to 
an overestimation of the reddening if it is determined from the 
Pa$\beta$/Pa$\gamma$ ratio or to an underestimation, if it is
derived from Br$\gamma$/Pa$\beta$. 

A previous determination of the reddening in Mrk~1210 was carried
out by Veilleux et al. (1997). They calculated a value of ${\rm E(B-V)}=0.48$ using the Balmer
decrement reported by Terlevich et al. (1991). Although this value
is smaller than the one determined here from the optical \ion{H}{i}, 
the discrepancy can likely be explained in terms of seeing, aperture and
position angle effects. Also, our result is slightly above the one
found in a recent statistical work presented by Ho et al.
(2003), which showed that Seyfert galaxies tend to have 
an internal mean reddening of about 0.43 (see also  Koski 1978).

The detection of the [\ion{S}{ii}] infrared lines near 1.03~$\mu$m and 
the optical lines at 4068,~4076~\AA, all
arising from the same upper $^2$P term, allows us to 
estimate the interstellar extinction affecting the S$^{+}$ region.  
This method was first pointed out by Miller (1968), and tested by 
Wampler (1968, 1971) in several Seyfert galaxies 
and more recently by 
Greve et al. (1994) in the Orion nebula. 
Due to the large separation in wavelength between the 
NIR and optical [\ion{S}{ii}], the reddening derived from this
ratio is, in principle, 
very accurate. 
Using the atomic data from Keenan (1991), we obtained a [\ion{S}{ii}] 
near-infrared to optical theoretical ratio of 0.67.
The observed ratio and the reddening coefficient are given in Table~\ref{redd}.
The E(B-V) of $ 0.28 \pm 0.08$ found for the S$^{+}$ region is the smallest one 
derived for the nuclear gas of Mrk~1210, 
indicating that these lines, as expected, are formed in the outer portions 
of the NLR.

In addition to the extinction determined from the \ion{H}{i} and
[\ion{S}{ii}] lines, the rich [\ion{Fe}{ii}] spectrum detected
in Mrk~1210 offers a unique opportunity to study the extinction
affecting the region where it is emitted. This is because
each of the three pairs of lines involved in the [\ion{Fe}{ii}] 
ratios listed in Table~\ref{redd} share 
the same upper level. It means that the intensity ratio depends only 
on energy differences between the lines and their Einstein A-coefficients,  
making these ratios useful probes of reddening. 
The scarce detection of such a rich [\ion{Fe}{ii}] spectrum  
in other AGN has prevented its use as diagnostics
of the internal extinction. For this reason, extinction 
estimates with the NIR [\ion{Fe}{ii}] lines are unusual 
and, to our knowledge, this is the first time that they are 
applied to a Seyfert nucleus (except for the ratio 
[\ion{Fe}{ii}]~1.256~$\mu$m/1.643~$\mu$m). Intrinsic line ratios
were taken from Bautista \& Pradhan (1998), calculated from  
Nussbaumer \& Storey (1988) transition probabilities. 

Our results, listed in Table~\ref{redd}, clearly show that 
large values of extinction affects the [\ion{Fe}{ii}] 
emission region. The fact that all three [\ion{Fe}{ii}] reddening
indicators agree within errors points out that all [\ion{Fe}{ii}]
is emitted in the same dusty environment. Although the 
uncertainties in some ratios are large, mostly because the lines
involved are intrinsically weak, the large extinction derived for the 
[\ion{Fe}{ii}] emitting region leads us to propose that the bulk 
of these lines is formed deep into the NLR, separate from the zone 
where other low-ionization species are emitted (i.e [\ion{S}{ii}] 
and [\ion{O}{i}]).

Contrasting to the nucleus, the reddening 
found for AP1 and AP3 show 
that the extended emission region is little or not affected
by dust (see Table~\ref{redd}).

The picture that emerges from the above analysis points
out that the dust distribution in the inner 250~pc of Mrk~1210
is rather inhomogeneous. Much of the optical 
emission we see should be emitted in the outer portions of the NLR 
whilst the NIR one comes from the deep NLR. This is supported
by the spectral narrow permitted and forbidden lines analysis made
in Sec.~\ref{kinematics}. At FHWM, the 
optical narrow components are narrower than the NIR ones. 
Moreover, the fact that the broad components display similar width, in velocity
space, from visible through NIR, suggests us that they are 
emitted in a region not affected by dust. 
Whether the large extinction measured from the [\ion{Fe}{ii}]
and the \ion{H}{i} lines is associated to the circumnuclear
starburst, to the host galaxy, or even to the torus cannot be
easily distinguished from our data. However, we should recall that
Martini et al. (2003), using HST images, classify Mrk~1210
as a tightly wound nuclear dust spiral with the individual dust
lanes traced over a full rotation about the nucleus. This view supports
the idea that the dust is mainly concentrated in a planar 
geometry, likely associated to the host galaxy. In this
scenario, the broad component arises in the outflow, which is
inclined, relative to our line of sight, from the dust plane.
Further support to this scenario comes from the dramatic
transition between a Compton-thick, reflection dominated state
and a Compton-thin state in Mrk~1210 reported by Guainazzi et
al. (2002) from X-ray observations. They claimed that this 
transition can be explained if the Compton-thick and 
Compton-thin absorbers are different. The former being associated
to the torus while the latter, located at larger scales,
may be associated with the host galaxy rather than with the
nuclear environment. 

\begin{table*}
\caption{Flux ratios and ${\rm E(B-V)}$.} \label{redd}

         \begin{center}

  \begin{tabular}[t]{cccccccc}
  \hline\hline
 & &\multicolumn{2}{c}{AP1}&\multicolumn{2}{c}{NUC}&\multicolumn{2}{c}{AP3}\\
  \cline{3-8}
 & Intrinsic & Measured  &  & 
Measured  &  & Measured & \\
Ratio & Value & Value & ${\rm E(B-V)}$ & Value & ${\rm E(B-V)}$ & Value &${\rm E(B-V)}$ \\
  \hline\hline
$\lbrack$\ion{Fe}{ii}$\rbrack$ 1.256/1.643\,$\mu$m & 1.36$^1$ & 
$1.42 \pm 0.18$ & $< 0.35$ & 
$0.89 \pm 0.04$ & $1.49 \pm 0.14$ &
$1.29 \pm 0.16$ & $< 0.62$ \\

$\lbrack$\ion{Fe}{ii}$\rbrack$ 1.676/1.294\,$\mu$m & 0.77$^1$ &
$2.50 \pm 1.34$ & $4.5 \pm 2.0$ & 
$1.14 \pm 0.14$ & $1.49 \pm 0.49$ &
$1.00 \pm 0.44$ & $< 2.72$ \\

$\lbrack$\ion{Fe}{ii}$\rbrack$ 1.533/1.297\,$\mu$m & 1.06$^1$ &
$3.0 \pm 12.0$ & $5.71 \pm 3.88$ & 
$1.27 \pm 0.07$ & $1.00 \pm 0.32$ &
-- & -- \\

Pa$\beta$/Pa$\gamma$& 1.79$^2$ & 
$1.92 \pm 0.33$ & $< 1.08$ & 
$2.24 \pm 0.14$ & $1.01 \pm 0.22$ & 
$1.89 \pm 0.54$ & $< 1.52$ \\

Br$\gamma$/Pa$\delta$& 0.50$^2$ & 
$0.50 \pm 0.14$ & $< 0.34$ & 
$0.84 \pm 0.10$ & $0.68 \pm 0.13$ & 
$0.60 \pm 0.35$ & $< 0.94$ \\

Br$\gamma$/Pa$\gamma$& 0.31$^2$ & 
$0.32 \pm 0.06$ & $< 0.35$ & 
$0.44 \pm 0.03$ & $0.52 \pm 0.10$ & 
$0.31 \pm 0.18$ & $< 0.88$ \\

Pa$\beta$/Pa$\delta$& 2.90$^2$ & 
$3.00 \pm 0.77$ & $< 0.80$ & 
$4.45 \pm 0.25$ & $1.15 \pm 0.25$ & 
$3.60 \pm 1.15$ & $< 1.45$ \\

Pa$\gamma$/Pa$\delta$& 1.62$^2$ & 
$1.56 \pm 0.46$ & $< 1.77$ & 
$1.98 \pm 0.21$ & $1.37 \pm 0.72$ & 
$1.89 \pm 0.76$ & $< 3.79$ \\

Br$\gamma$/Pa$\beta$ & 0.17$^2$ & 
$0.17 \pm 0.02$ & $< 0.27$ & 
$0.19 \pm 0.01$ & $0.31 \pm 0.10$ & 
$0.17 \pm 0.08$ & $< 1.12$ \\

  \hline
H$\alpha$/H$\beta$ & 3.10$^2$ & & & $ 5.77 \pm 0.27$ & $ 0.62 \pm 0.05$ & \\
$\lbrack$\ion{S}{ii}$\rbrack$ 1.03\,$\mu$m/$4068\,\AA$ & 0.67$^3$ &&& $1.55 \pm 0.39$ & 
$0.28 \pm 0.08$ & \\
\hline
\hline
   \end{tabular}
\end{center}
$^1$ Calculated from  Nussbaumer 
               \& Storey (1988) transition probabilities.\\
$^2$ Osterbrock (1989).\\
$^3$ Calculated from Keenan (1991) atomic data. 
\end{table*}


\section{Electron Density and Temperature: physical properties of 
the nuclear environment} \label{physprop}

Electronic densities and temperatures for the nuclear gas of Mrk~1210 
can be determined by means of several diagnostic line ratios. 
But before this can be done, line fluxes need to be corrected in 
accord to the value of reddening obtained 
from lines of the same ion or similar ions. Thus,
low-ionization lines such as [\ion{O}{i}], [\ion{N}{i}], [\ion{N}{ii}] and 
[\ion{S}{ii}]
were corrected by $ {\rm E(B-V)}=0.28$. This value was determined
from the [\ion{S}{ii}] lines. The exception of this low-ionization
group is [\ion{Fe}{ii}], whose lines were corrected by ${\rm E(B-V)}=1.5$, 
as measured from the different indicators employed. 
The  remaining lines were corrected 
for ${\rm E(B-V)}=0.5$. It results from 
averaging out our determination of internal extinction by means of optical 
\ion{H}{i} lines, the one reported by Veilleux et al. (1997) for Mrk~1210
and the average internal extinction reported by Ho et al. (2003).


In the low-density regime, the [\ion{S}{ii}]~$\lambda$6716/$\lambda$6730 
flux ratio is a well known diagnostics of the
electronic density that is relatively insensitive to temperature. 
This diagnostic can be applied to regions where the electronic density is
smaller 
than the critical density of the [\ion{S}{ii}] lines 
($n_\mathrm{c}\sim 10^{4}$~cm$^{-3}$). Above this density the lines become 
collisionally de-excited. The observed [\ion{S}{ii}] ratio of 0.9
indicates an electron density $n_\mathrm{e} \sim 1\,000$~cm$^{-3}$ for 
temperatures in the interval 10\,000~K to 50\,000~K.
On the other hand, the infrared [\ion{S}{ii}]~1.03~$\mu$m lines are 
observed to 
be quite strong relatively to the 6716, 6730~\AA\ doublet 
in the nuclei of Mrk~1210, indicating that a second, 
higher density ($n_\mathrm{e} \geq 10^{4}$~cm$^{-3}$) S$^{+}$ zone 
can also exist. The [\ion{S}{ii}]~1.03~\micron/(\l6716+\l6730) line ratio is 
very sensitive to variations in density, in particular for values of 
$n_\mathrm{e} \geq 10^{4}$~cm$^{-3}$ where the optical [\ion{S}{ii}] 
diagnostic no longer varies with $n_\mathrm{e}$.
The observed 
[\ion{S}{ii}]~1.03~\micron/(\l6716+\l6730) line ratio is 
0.59, which indicates an electron density $\sim 13\,000$~cm$^{-3}$ 
for an assumed temperature T$_\mathrm{e}=10\,000$~K (Keenan 1991). 
The differences between the densities derived from the two
[\ion{S}{ii}] line flux 
ratios illustrate that the [\ion{S}{ii}] lines observed in the spectrum 
of the central region of Mrk~1210 arise from different zones, with 
different densities: a denser region ($\sim 13\,000$~cm$^{-3}$) 
that produces the bulk of the 
[\ion{S}{ii}] \l4068, \l4076\ and 1.03~\micron\ emission and a lower density 
region ($\sim 1\,000$~cm$^{-3}$) that contributes with most of the 
\l6716 and \l6730 flux.

The density of the Fe$^+$ emitting region can be derived from the observed 
\fe2~1.533~\micron/1.643~\micron\ and \fe2~0.861~\micron/1.256~\micron\ line 
ratios (Pradhan \& Zhang 1993). The former ratio yields a density
$n_\mathrm{e}\sim10^{4.2}$~cm$^{-3}$ for T$_\mathrm{e}$ in the
interval 3\,000 -- 12\,000~K, while from the latter ratio we derive 
a higher density, $n_\mathrm{e}\sim10^{5.2}$~cm$^{-3}$, assuming
T$_\mathrm{e}=10\,000$~K. These results, together with the ones from the 
previous sections, strongly suggest that the \fe2\ lines are indeed 
emitted in a separate 
region, characterized with higher densities and higher extinction.

\begin{figure}
  \centering
  \includegraphics[width=9cm]{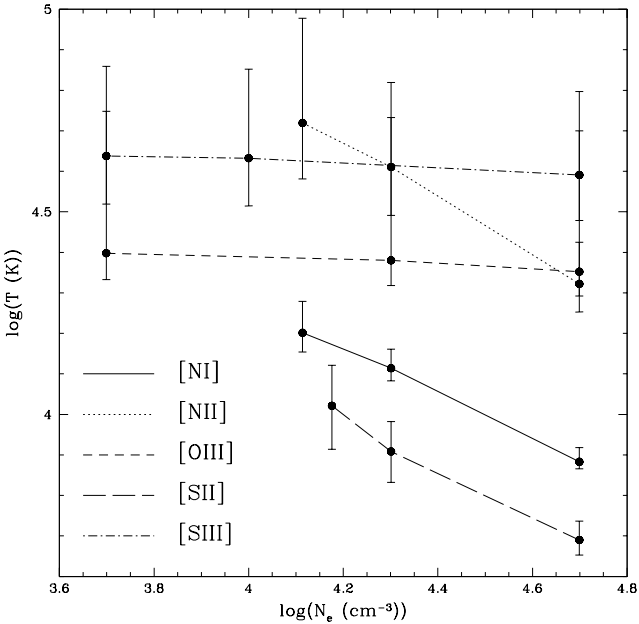}
  \caption{Electron densities and temperatures 
found for Mrk~1210.}\label{temp}
\end{figure}

The gas temperature can be estimated from measurements of pairs of 
emission line ratios emitted by a single ion from two levels 
with considerably different excitation energies. Two good examples 
are the [\ion{O}{iii}]~(\l4958+\l5006)/\l4363 and 
[\ion{N}{ii}]~(\l6548+\l6583)/\l5754 
line ratios. Temperature determinations from these ratios were calculated, 
using the TEMDEN task of IRAF STSDAS nebular package. The results obtained 
for a wide range 
of densities are plotted in Figure~\ref{temp}, which shows very high 
temperatures in the O$^{+2}$ and N$^{+}$ zones, ranging from 18\,000~K 
up to 50\,000~K.
Although the ionization potential required to form the O$^{+2}$ and 
N$^{+}$ ions (35.1 and 14.5~eV, respectively) are different, the 
critical densities for the upper levels of the 
lines involved in the 
[\ion{O}{iii}] and [\ion{N}{ii}] line ratios are $\geq 10^{4.9}$~cm$^{-3}$. 
Thus it is reasonable to suppose that these lines arise from 
the same region with $n_\mathrm{e} \leq n_\mathrm{c}$. 
Therefore, we set a common temperature of T$_\mathrm{e}=22\,000$~K, 
for an assumed density $n_\mathrm{e}=50\,000$~cm$^{-3}$. 
Electron temperatures T$_\mathrm{e}>20\,000$~K are difficult to obtain 
in the O$^{+2}$ and N$^{+}$ zones in photoionized gas, 
requiring a source of energy 
input in addition to photoionization. Our working hypothesis
is that this source is associated to shock heating.

Another well known temperature diagnostic is the 
[\ion{S}{iii}]~(0.906+0.953)~\micron/\l6312 line ratio, which is analogous 
to the [\ion{O}{iii}]~(\l4958+\l5006)/\l4363 ratio. Nonetheless, it
is not commonly used, mainly due to the problem of deblending 
[\ion{S}{iii}]~\l6312 from the usually considerably stronger 
[\ion{O}{i}]~\l6300. Our medium resolution optical spectrum allowed us
to separate these two lines without introducing large uncertainties
in the line fluxes. 
From the [\ion{S}{iii}], we 
calculated the temperature of the S$^{+2}$ region using the TEMDEN task 
included in the STSDAS Version 3.3
package of IRAF. The atomic data employed by the routine are at least as
recent as those given in the compilation by Pradhan and Peng (1995). The results are also plotted in Figure~\ref{temp}. Note that the 
calculation is relatively insensitive to density, indicating a 
gas temperature of about 40\,000~K for densities in the interval 
10$^4-5\times10^4$~cm$^{-3}$. 
Osterbrock et al. (1990) determined the S$^{+2}$ gas temperature 
for several Seyfert galaxies, finding temperatures from 10\,000~K to 
the extreme case of 26\,100~K for the high-ionization Seyfert~1 
galaxy III Zw 77. The unusually high temperature found for the S$^{+2}$ 
emitting region of Mrk~1210, together with the high temperature 
derived from the [\ion{O}{iii}] line ratio, shows further 
evidence that shocks contribute 
to the ionization structure in the NLR of this galaxy. 
Recall that these
lines are the ones with a blueshifted broad component that we
associated to the radio-knots observed in VLBA maps.

Also plotted in Fig.~\ref{temp} are the electron temperature 
in the N$^{0}$ and S$^{+}$ zones, derived from 
the temperature sensitive [\ion{N}{i}]~\l5197/1.040~\micron\ and 
[\ion{S}{ii}]~(\l6716+\l6730)/\l4068 line ratios, respectively. 

The ratio [\ion{O}{i}]~(\l6300+\l6363)/\l5577 also works 
as a temperature indicator. 
Although the line [\ion{O}{i}]~\l5577 
is definitely present in the optical spectrum of Mrk~1210, 
the profile is too noisy to measure it accurately. Therefore, we could only 
set an upper limit for the line flux, which we used to derive an upper limit 
of the [\ion{O}{i}] emitting gas temperature of 12\,000~K. The low temperature 
of the gas, 
in addition to the low ionization potential of the ion indicates that 
these lines are probably emitted in the same region as 
[\ion{S}{ii}] and [\ion{N}{i}]. 

The results found from the different indicators (see Table~\ref{temden}) 
show that the NLR of Mrk~1210
cannot be characterized by a single value of temperature and density, but that 
presents a noticeable gradient in the physical conditions. 
The temperatures range from about 10\,000~K up to the unusual value of 
40\,000~K determined from the [\ion{S}{iii}] lines, although this latter
value may be misleading because of the contribution from shocks. 
The range of densities 
extends at least over three orders of magnitude, from a low density 
region ($n_\mathrm{e}=1\,000$~cm$^{-3}$) in which the 
[\ion{S}{ii}]~\l6716, \l6730 lines (and probably the [\ion{N}{i}] and 
[\ion{O}{i}] lines) are emitted, to a high density region 
($n_\mathrm{e}=10^{5.2}$~cm$^{-3}$) where the 
[\ion{Fe}{ii}] lines are emitted. Note that densities that high are
plausible for the [\ion{Fe}{ii}] given that the critical densities of most
if its transitions are in the interval 10$^4$--10$^5$~cm$^{-3}$ 
(Nisini et al. 2002).

\begin{table*}
\caption{Density and temperatures found from different diagnostic line
ratios.}\label{temden}
\begin{center}
\begin{tabular}[t]{lcc}
\hline\hline
Density sensitive line ratio & Assumed temperature [K] & Density [cm$^{-3}$]\\
\hline
$\lbrack$\ion{S}{ii}$\rbrack$ $6716/6730~\AA$ & $10\,000-50\,000$ & $\sim 10^{3.0}$ \\
$\lbrack$\ion{S}{ii}$\rbrack$ 1.03~$\mu$m/$(6716+6730)~\AA$ & 10\,000 & $ \sim 10^{4.1}$ \\
$\lbrack$\ion{Fe}{ii}$\rbrack$ 1.533/1.643~$\mu$m & $3\,000-12\,000$ & $\sim 10^{4.2}$ \\
$\lbrack$\ion{Fe}{ii}$\rbrack$ 0.861/1.256~$\mu$m & 10\,000 & $\sim 10^{5.2}$\\
\hline\hline
Temperature sensitive line ratio & Assumed density [cm$^{-3}$]& Temperature [K]\\
\hline
$\lbrack$\ion{N}{i}$\rbrack$ 5197~\AA/1.040~$\mu$m & $10^{4.1}-10^{4.7}$ & $7\,000-16\,000$ \\
$\lbrack$\ion{N}{ii}$\rbrack$ $(6548+6583)/5754~\AA$ & $10^{4.1}-10^{4.7}$
&$21\,000-52\,000$ \\
$\lbrack$\ion{O}{i}$\rbrack$ $(6300+6363)/5577~\AA$ & $10^{3.7}-10^{4.7}$ & $< 12\,000$\\
$\lbrack$\ion{O}{iii}$\rbrack$ $(4958+5006)/4363~\AA$ &$10^{3.7}-10^{4.7}$ &
$22\,500-25\,000$\\
$\lbrack$\ion{S}{ii}$\rbrack$ $(6716+6730)/4068~\AA$ &
$10^{4.1}-10^{4.7}$ & $5\,000-12\,500$ \\
$\lbrack$\ion{S}{iii}$\rbrack$ $(0.906+0.953)~\mu$m/6312~\AA & $10^{3.7}-10^{4.7}$ & $39\,000-43\,500$\\
\hline\hline
\end{tabular}
\end{center}
\end{table*}

\section{Summary and Conclusions} \label{conclusions}
We have carried out a detailed study of the physical properties of the nuclear 
and extended emission line regions of the Seyfert~2 galaxy Mrk~1210. 
For this purpose, we presented spectroscopic observations 
covering simultaneously the interval 0.8--2.4~$\mu$m. 
The spectrum in the range 0.81--1.03~$\mu$m is the first one published for this
object. For that reason, all lines detected in this region were not previously 
identified in this source. 

The spectra of Mrk~1210 display a plethora
of emission lines, from those emitted from molecular H$_{2}$, prominent 
mostly in
the K-band, along with low and medium ionization lines such as [\ion{C}{i}],
[\ion{S}{ii}] and [\ion{S}{iii}] and up to very high ionization lines of 
[\ion{S}{ix}] and [\ion{Si}{x}]. In addition, Mrk~1210 is characterized by a 
rich [\ion{Fe}{ii}] spectrum, not previously reported in the literature for 
other AGN. Extended emission of [\ion{S}{iii}], [\ion{Fe}{ii}] and
\ion{He}{i} is found up to a distance of 500~pc from the nucleus.

The large amount of spectral information, at medium resolution, 
covering the optical and NIR regions allow us to study the physical conditions
in the inner few hundred parsecs of this galaxy. Our main results can be
summarized as follows:

{\it (i)} From the depth of the $^{12}{\rm CO(6-3)}$ overtone bandhead 
at 1.618~$\mu$m, we have estimated that $83\% \pm 8\%$  of the H-band 
continuum is of stellar origin. This value improves previous estimates,
which claimed that at least 50\% of the H-band continuum was attributed
to the AGN and agrees with results found from optical spectroscopy.
It also suggests the usefulness of the NIR in determining the percentage
contribution of the stellar population to the integrated continuum 
emission. After correcting for the increased contribution of the 
stellar population to the continuum, Mrk~1210 continues displaying
one of the highest star formation efficiency
among water megamaser galaxies in spite of its relatively low
molecular gas content. This result provides further support to previous
findings showing that the molecular lines in Mrk~1210 were mainly excited
by UV heating from stars.

{\it (ii)} The analysis of the emission line profiles, both allowed and 
forbidden, shows that most prominent lines are characterized by a narrow 
(${\rm FWHM} \sim 500$~\kms) component on top of a broad 
(${\rm FWHM} > 1\,000$~\kms) 
blue-shifted component ($\Delta{\rm V} \sim 150$~\kms). The latter seems 
to be associated to a nuclear outflow, instead of the hidden 
BLR claimed to be present in previous NIR observations of this object.
Moreover, the  differences in form and width of the line profiles of
[\ion{S}{iii}], [\ion{Fe}{ii}] and H$_2$ imply that the 
emission seen from each line originates in different volumes of gas.

{\it (iii)} We have examined the internal extinction affecting the NLR within 
the inner few hundred parsec using several indicators, including 
[\ion{S}{ii}] and [\ion{Fe}{ii}] line ratios. The results reveal a dusty AGN 
while the extended regions are very little or not affected 
by dust. In the inner 250~pc we found an 
${\rm E(B-V)}=0.5$ from the \ion{H}{i} lines ratios. 
From the [\ion{Fe}{ii}] line ratios, a larger extinction 
value, ${\rm E(B-V)}=1.5$, was derived. This supports the hypothesis that the 
[\ion{Fe}{ii}] lines are formed in a separate region, different from the 
partially ionized zone that exist in AGN. At 250~pc from the center, no 
extinction is found from our data.

{\it (iv)} Electronic densities and temperatures for the nuclear gas of Mrk~1210 
were determined by means of several diagnostic line ratios. The results found 
from the different indicators show that the NLR of Mrk~1210
cannot be characterized by a single temperature and density, since it  
presents a noticeable gradient in the physical conditions. 
The temperatures are from about 10\,000~K up to the unusual value of 
40\,000~K determined from the [\ion{S}{iii}] lines. This very high 
temperature, 
together with the high temperature derived from the [\ion{O}{iii}] line ratio and 
the fact that these two lines display broad blueshifted components 
are evidence that shocks contribute to the ionization structure of the NLR of this galaxy. 
The range of NLR gas densities extends over three orders of magnitude 
and points out to a low density 
region ($n_\mathrm{e}=1\,000$~cm$^{-3}$) in which the 
[\ion{S}{ii}]~\l6716, \l6730 lines (and probably the [\ion{N}{i}] and 
[\ion{O}{i}] lines) are emitted, up to a high density region 
($n_\mathrm{e}=10^{5.2}$~cm$^{-3}$) where the 
[\ion{Fe}{ii}] lines are emitted.

\acknowledgements

This research has been partly supported by the Brazilian agency
CNPq (309054/03-6) to ARA and the European 
Commission's ALFA-II program through its funding of the Latin-American 
European Network for Astrophysics and Cosmology, LENAC to XM. 
The authors thank to the Referee Enrique P\'erez for its
useful comments to improve this manuscript. This research has 
made use of the NASA/IPAC Extragalactic Database (NED) which is 
operated by the Jet Propulsion Laboratory, California Institute of 
Technology, under contract with the National Aeronautics and Space 
Administration.

\end{document}